\documentclass[12pt]{article}

\usepackage{scicite}

\usepackage{times}

\usepackage{graphicx}

\newenvironment{sciabstract}{%
\begin{quote} \bf}
{\end{quote}}

\title{Comprehensive Investigation and Isolation have Effectively Suppressed the Spread of COVID-19}

\author
{Yubo Huang$^{1\ast}$, Weidong Zhang$^1\ast$\\
\\
\normalsize{$^{1}$Department of Automation, Shanghai Jiao Tong university, Shanghai 200240, China,}\\
\normalsize{$^\ast$Corresponding author; E-mail: huangyubo@sjtu.edu.cn, wdzhang@sjtu.edu.cn}
}

\date{}

\begin{document}


\baselineskip24pt


\maketitle


\begin{sciabstract}
The outbreak of COVID-19 since Dec. 2019 has caused severe life and economic damage worldwide, many countries are trapped by medical resource constraints or absence of targeted therapeutics, and therefore the implement of systemic policies to block this pandemic should be prioritized. Based on the transmission mechanisms and physicochemical properties of betacoronaviruses, we construct a fine-grained transmission dynamics model (ICRD) to forecast the crucial information of public concern, therein using dynamical coefficients to quantify the impact of the implement time and intensity of containment policies on the spread of epidemic. We find that the comprehensive investigation policy for susceptible population and the quarantine for suspected cases eminently contribute to reduce casualties during the phase of the dramatic increase of diagnosed cases. Statistic evidences strongly suggest that society should take such forceful public health interventions to cut the infection channels in the initial stage until the pandemic is interrupted.
\end{sciabstract}


\section*{Introduction}
Pneumonia caused by COVID-19 has evolved a worldwide pandemic for which humans have paid more than 2.5 million infections and 180 thousand deaths as of Apr. 21th, 2020~\cite{xu2020epidemiological} but there are still billions of lives threatened~\cite{world2020coronavirus,wolfel2020virological,shi2020radiological}. The transmission of the COVID-19 virus is confirmed through respiratory droplet\cite{bourouiba2020turbulent} and has super infectivity beyond SARS and MERS~\cite{peeri2020sars} which can even overwhelm countries with advanced medical facilities and therapeutics. Laboratory experiments proved that SARS-CoV-2 can survive temperatures of 56 $^\circ C$ for more than 30 minutes~\cite{pastorino2020evaluation} and the experiences from tropical regions~\cite{sajadi2020temperature} illustrated COVID-19 still has transmission capacity under the hydrothermal circumstance~\cite{luo2020possible}. Moreover, the incubation period of COVID-19 is relatively short (5.5 days average) implying that it quickly onset after infection and presents the corresponding clinical symptoms such as fever, cough, and fatigue~\cite{lauer2020incubation}. Recently, the revised data released by Chinese health authorities definitely delivered that the case fatality rate (CFI) of COVID-19 is seriously underestimated because the phenomenon of missed reports and false positive in the early stage of the epidemic. Given the current statue of the limited pharmaceutical treatments and the unavailable targeted vaccines~\cite{lurie2020developing}, the public health interventions are demonstrated as the most functional precautions in many countries~\cite{yang2020combating,hartley2020public,hellewell2020feasibility}, especially contributing to block the pandemic in Wuhan\cite{kraemer2020effect,pan2020association,tian2020investigation,prem2020effect}.

On Jan. 23th 2020, Chinese authorities imposed a lockdown in Wuhan since the suspected and diagnosed cases exceed 2,000 which triggers public health emergency response in Hubei Province. The containment efforts including school closures, transports bans and workplace shutdowns were announced to limit the spread of virus. Nevertheless, the epidemic entered the exponential period unstoppably from Feb. onwards~\cite{li2020early,wu2020estimating} and caused global concern and panic. From Feb. 2nd, the successive delivery of the shelter hospitals implied Wuhan had the sufficient wards to isolate and treat symptomatic patients. Meanwhile, people strictly followed the ordinances concerning social distancing, quarantine and careful hand and respiratory hygiene, subsequently wearing personal protective equipments (PPE)~\cite{leung2020respiratory}, isolating themselves at home, and recording their temperature each day. The communities adopted a comprehensive investigation strategy to detect the suspected cases and reported their trace while hiding their private information~\cite{ferretti2020quantifying} resulting in a sub-exponential growth of the infected curve until a peak of 38,020 on Feb. 18th~\cite{maier2020effective}. Despite there is a ethical controversy about whether suspicious persons should be mandatorily isolated from social networks, the experience from Wuhan supported that this policy had effectively cut the transmission channels of COVID-19 and compel the pandemic to a mitigation stage. In post-pandemic phase~\cite{kissler2020projecting}, Wuhan continued to distribute disinfectors throughout the region including ventilation duct, water pipe, and urban sewers where COVID-19 was sampled in Paris~\cite{wurtzer2020time} and the central China government stated the pandemic was basically blocked on Mar. 23th. Ultimately, after paying 4,632 deaths nationwide and incalculable economic cost, China has ended its 76-day lockdown of Wuhan, but the relative restrictions still remain in place.

The transmission dynamics observed in different continents and zones are markedly heterogeneous consulting time series data released by public health authorities\cite{gudbjartsson2020spread,fanelli2020analysis,tsang2020impact,salathe2020covid}. The fundamental reasons are that the fluctuation of the objective physical environments and the implementation intensity and duration of the containments policies restricted by public opinion, both presenting challenges to build the dynamic model to forecast the tendency of the pandemic. Furthermore, the CFI and recovery rate change as well with respect to different phases since the gradual enrichment of care experiences and capacities. Therefore, we design dynamical coefficients to quantify the variation of infectivity, investigation and isolation policies, vital dynamics to adjust the social response to the pandemic. Functionally, tuning the boundaries or the derived functions of the dynamical coefficients can subtly regulate the containment efforts and further observe their impact on the infected curves. We then simulate the results about advancing or postponing relative policies and conclude that the output curves are sensitive to these policies. The results of our model introduce statistic evidences that the containment policies can effectively suppress or even block the outbreak of COVID-19 through mapping them into measurable interval coefficients to observe their influence on the epidemic.

\section*{Modeling transmission dynamics of COVID-19}
From the cases and traces information of patients released by the Wuhan government and CDC, most individuals were infected from a symptomatic or pre-symptomatic infection, especially during the period of exponential and sub-exponential growth, whereas the persons who were infected through asymptomatic or environmental transmission merely accounted for a negligible fraction of infected population. In practical, the boundaries of these four transmissions are ambiguous and the number of asymptomatic infections and secondary infections caused by them is difficult to count. Therefore, we generally assume that the non-isolated infected individuals are the main propagating sources on social networks. We divide the statues of the population into 5 categories: (H)ealthy, (I)nfected, (C)onfirmed, (R)ecovered and (D)ead (Fig. 1). More granular, the infected group $I$ consists of the confirmed $C$ and unconfirmed $I-C$ and therein the confirmed individuals $C$ are either isolated $\beta C$ ($\beta$ is the isolation rate) or non-isolated $(1 - \beta)C$. The transmission dynamics of virus can be fully described by the ordinary differential equations (namely ICRD model) with respect to time $t$:
\begin{eqnarray}
 \partial_t I &=& \alpha (I - \beta C) (1-(\frac{I}{N})^{-\gamma}) - (\sigma + \kappa)C - (\delta + \mu) (I - C) \\
 \partial_t C &=& \eta (I - C) - \sigma C - \kappa C \\
 \partial_t R &=& \sigma C + \delta(I - C) \\
 \partial_t D &=& \kappa C + \mu(I - C)
\end{eqnarray}
The infected but non-isolated group $I - \beta C$ has the vital infectiousness since they not only are the virus carriers but can randomly walk on social networks. The average number of secondary infections an infected would induce over the period from infected to isolated $T$ (transmission is terminated) is defined as the basic reproduction number $R_0$. The infectiousness intensity in a unit period is $\alpha = R_0 / T$ and the new increase of infected individuals thereby are $\alpha (I - \beta C)$. Apparently, $\alpha$ is inversely proportional to social distancing, determining the increment of infected cases each day and tuning $\alpha$ could capture the effect of national isolation policies subtly. Nevertheless, one person would be repeatedly secondarily infected with high probability when the infection density in the community reaches a critical level and above statistic value significantly greater than the true value. We then present a harmonic function $(1-(I/N)^{-\gamma})$ to tackle the redundancy counting dilemma, in which $N$ represents the community population and $\gamma$ denotes the harmonic coefficient. $\alpha (1-(I/N)^{-\gamma})$ shows the infectivity of an individual would degrade as the group density of infection increases. Meanwhile, there would be a proportion of infected people who recover (cured after treatment $\sigma C$ or spontaneous recovery $\kappa C$) or die (die with treatment $\delta (I-C)$ or without treatment $\mu (I-C)$) in one unit statistical period, where $\sigma, \kappa, \delta, \mu$ are daily cure rate, incurable mortality rate, natural recovery rate, and non-treatment mortality rate, respectively. Hence $\partial_t I$ quantifies the aggregate incremental cases of infection (Eq. (1)) after removing the recovered and dead population. Limiting to the objective medical testing ability of laboratories, only a fraction of $\eta$ suspected cases could receive diagnosis within the unit statistical time ($\eta$ is the algebraic mapping of investigation policy), so the increment of confirmed cases $\partial_t C$ equals to the difference between the newly positive diagnosed cases $\eta(I-C)$ and dead or recovered cases $(\sigma + \kappa)C$. Evently, we could deduce the daily increase of recovered or dead cases integrating the recovery ratio $\sigma, \delta$ and mortality $\kappa, \mu$ with the infected population (please see Materials and Methods for detailed derivation of Eqs. (1-4)).

\section*{Using dynamical parameters to quantify the investigation and isolation policies}
Before we have discussed that $\beta$ and $\eta$ reflect the impact of the isolation and investigation policies on a pandemic in ICRD model, respectively. The constant coefficients in Eq. (1-4) assume that the containment efforts or the physical and chemical characteristics of the coronavirus remain permanent in long-term transmission, and undoubtedly it is counterintuitive. In fact, the governments tend to adopt more compulsive policies as the dramatic increase of death cases, residents choose isolate themselves spontaneously with more publicity on negative news, and the testing and diagnosis techniques would be visibly improved after the deeply studying of the virus's properties. Hence the measures and intensity of the society to fight the pandemic change in different phases. For instance, initially, due to the lack of awareness of the virus, people are reluctant to maintain social distancing with others ($R_0$ and $\alpha$ are relatively large), so the virus has explosive infectivity at this stage. Subsequently, with the exponential increase of infected cases, the government would announce voluntary or compulsive isolation policies to reduce $\alpha$ until it remains low (Fig. 2\textbf{a}). Furthermore, many susceptible patients cannot be diagnosed because of the limited medical resource in the preliminary stage. However, the promotion of the testing techniques and the enhancement of the investigation policies ($\eta$ rises) would strongly reverse this situation in the successive stages. Therefore, we introduce the dynamic coefficients to simulate the transformation of propagation:
\begin{eqnarray}
\alpha &=& a * e^{-bt} + c \\
\eta &=& u * \arctan(vt) + w
\end{eqnarray}
Fig. 2 illustrates the curves of dynamical $\alpha$ and $\eta$, and it is clear that the parameters $a, c$ (or $u,v$) determine the boundaries of $\alpha$ (or $\eta$) that catch intensity scope of the isolation (investigation) policies and $b$ (or $v$) is a critic to evaluate the response time of these policies, both regulating the tendency of epidemic (see Supply material for other dynamic coefficients and the tuning methods).

\section*{Statistic evidences for the power of comprehensive investigation and compulsive isolation policies}
Fig. 3\textbf{A} shows that SEIR model forecasts the existing infected cases had reached a peak of 4,885,672 at Mar. 3rd that two orders of magnitude higher than the actual value and the transmission would be effectively controlled before Apr. 17th without external constraints. SEIR inappropriately concludes that more than 50\% of the population in Wuhan would be infected in this worldwide epidemic, and unfortunately it ignores the dead cases caused by COVID-19. Static coefficients based-ICRD (S-ICRD) model predicts the existing infected cases would sub-exponentially increase to 4,2729 by Feb. 8th and the pandemic would be basically blocked at May. 27th. It is noticeable that the infected curve in Fig. 3\textbf{B} declines extraordinarily slower than the clinical statistics resulting from the illogical assumption of constant public health interventions and it explains why the eventual death toll (8,195) markedly higher than the official surveys (3,869) as well.

Compared with the SEIR and S-ICRD, the dynamic coefficients based-ICRD model (D-ICRD) has infinitesimal mean square errors (MSE) with the data released by NHC (Fig. 4). D-ICRD estimates that the existing infected and confirmed cases would dramatically raised to their peaks (41986, 38758) at Feb. 14th and 17th respectively (Fig 4\textbf{A}), approximately consisting with the practical inflexion Feb. 18th (38020). In Phase 1, the incomplete investigation policies and limited testing techniques contribute the large interval between the infected and confirmed curves, whereas they gradually coincide with the strengthening of surveillance in phase 2. The truth-values of the recovered cases are always located in the 95\% confidence interval (CI) of the curve predicted by D-ICRD but the estimated values are slight higher since the patients who are spontaneously recovered were excluded in the statistics (Fig. 4\textbf{B}). The death cases calculated by D-ICRD eccentrically deviate from the statistical data before Mar. 8th in Fig. 4\textbf{C}. Until Apr. 17th the government announced that there were 1290 missing victims in the initial collecting stage and then increased death cases to 3869 that located in the 95\% CI [3576, 3952] of the predicted value.

Fig. 2 straightway illustrates that controlling $b$ (or $v$) can accurately regulate the implement time of the isolation (investigation) policies and the corresponding transmission results are shown in Fig. 5. The peaks of the existing infected cases would reduced 12,725 or 17,009 if the isolation policies were executed 2 or 3 days in advance respectively, and thousands of innocent individuals would been saved (Fig. 5\textbf{B}). Inversely, postponing the quarantine policies for 4 (or 7) days would induce another 15,900 (or 53,753) people to be infected and increase the death toll to 5,454 (or 6,783). The advancement or delay of the investigation policies would also cause similar results, but their intensity is softer than the isolation policies. For instance, the comprehensive investigation order issued 7 or 12 days earlier would protect 9,567 or 15, 709 people from infection, respectively. Meanwhile, benefiting from the quarantine of suspected persons, medical pressure would be relieved and then 661 or 940 sufferers would be rescued from death. Nevertheless, if such policies were delayed for 17 days, more than a quarter of contacts would be infected or die considering that the increasing of suspected nodes on social networks incredibly shorten their connections with the healthy nodes. Accordingly, the above evidences definitely exhibit that this pandemic is extremely sensitive to the public health interventions, particularly in the initial stage, which can forcefully cut off the possible transmission routes of COVID-19. Beyond that, D-ICRD also confirms that measures, such as using personal protective equipments (PPE, lowering $\alpha$), monitoring and recording temperature (increasing $\sigma$), declaring community emergency response plan (reducing $\mu$) and so on, can all effectively suppress the spread of virus.

\section*{Discussion}
Due to the Wuhan government implemented mandatory closures at the early stage of the epidemic to prohibit the free movement of population and issued home quarantine order to isolate internal social contacts during the Spring festival, Wuhan is an ideal research object about the transmission of COVID-19, in which we can detect the general propagation laws about this pandemic. Based on the transmission mechanisms and physicochemical properties of coronavirus, in this paper, we builded a dynamics model named D-ICRD to observe the dynamical change of the existing infected cases, the existing confirmed cases, the cumulative recovered cases and the cumulative dead cases with respect to time in Wuhan. Beyond that, our model could also deduce the significant pandemic indexes such that cumulative infected cases, cumulative confirmed cases, new suspected cases etc. Functionally, D-ICRD could not only display the characteristic of nature virus transmission but also precisely quantify the influence of the extra containments on the spread of COVID-19.

We designed the dynamical coefficients associating with the implement time and intensity of the relative policies to simulate the transmission environment of COVID-19 in Wuhan. Compared with the static coefficients, the dynamical coefficients can more penetratingly capture the rational adjustment of social response to the severity of the epidemic. For example, many classical models assume the constant basic reproduction number but we believe that it changes with the intervention factors and eventually this change is mapped on the transmission coefficient $\alpha$. Moreover, the improvement of therapy and the use of specific medicines would significantly affect the cure rate and mortality. Hereupon D-ICRD entirely reproduced the time series data released by NHC, and it balances robustness and agility.

Then, we subtly regulated the implement time or intensity of these interventional policies through changing the derived functions or boundaries of the corresponding dynamical coefficients to verify the contribution of such policies to suppress the epidemic. We found that the infected and death cases would sharply decrease if the quarantine strategies could be promulgated slightly in advance. Nevertheless, D-ICRD indicates postponing such plans would cause at least additional millions of infected population worldwide, and undoubtedly the catastrophic consequences of abrogating these ordinances are uncountable. Furthermore, the intensity of the isolation and investigation polices would determine the trend of the pandemic as well. The statistical evidences show the forceful public health interventions can effectively suppress or even block the outbreak based on Wuhan's experiences although COVID-19 has extremely strong infectivity (droplet transmission and suspected aerosol transmission) and lethality (CFR $>$ 7.69\% in Wuhan, before Feb. 20th). We suggest that the governments or communities should adopt the containments such as comprehensive investigating and tracking the temperature and traces information about the residents, immediately isolating the suspected and diagnosed infected individuals to prevent the spread of COVID-19 on social networks. Although such near-compulsory national policies may cause severe economic recession or ethical problems, we must realize that society is a community of life composed of all human beings, and we profoundly influence each other whereupon we should collaborate to fight this vital pandemic together. Given that there are no target therapeutics or vaccine currently, each of us should cooperate with these policies in a responsible manner, isolate ourselves from others, form an island, and ultimately prevent the spread of the virus.

\bibliography{scibib}

\bibliographystyle{Science}

\section*{Acknowledgments}
Thanks all health care workers worldwide for their commitment, dedication, and professionalism in COVID-19 pandemic. This paper is partly supported by the National Science Foundation of China (61473183, U1509211, 61627810), National Key R$\&$D Program of China (SQ2017YFGH001005). No conflicts of interest, financial or otherwise, are declared by the authors.
\section*{Supplementary materials}
Materials and Methods\\
Supplementary Text\\
Figs. S1 to S3\\
Tables S1 to S4\\
References \textit{(4-10)}


\clearpage


\begin{figure}
  \centering
  \includegraphics[width=17cm]{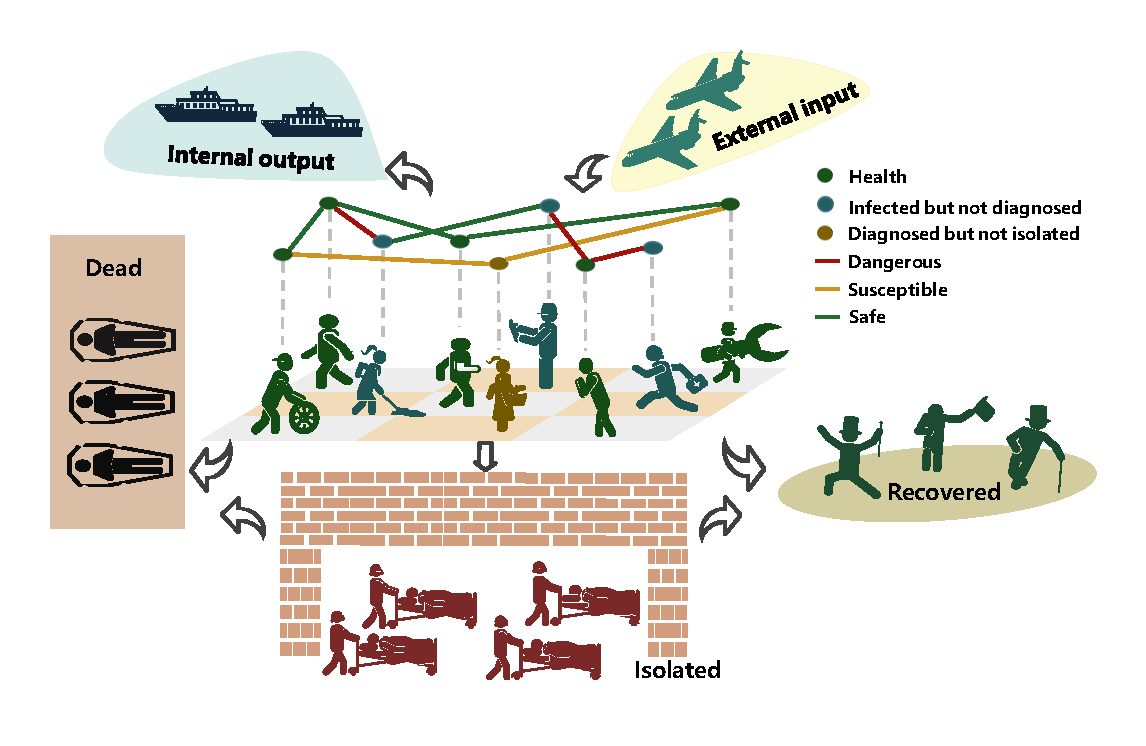}
  \caption{\textbf{Transmission mechanism of COVID-19 on social networks.} The group of infected but non-isolated (blue and yellow nodes) is the main propagating source of virus through physical contacts (droplet), especially over the incubation period~\cite{kucharski2020early,li2020substantial}. The infected patients would die or produce antibodies without treatment but undoubtedly the herd immunity would take a long time to achieve and uncountable individuals would lose their lives in this process. Therefore, the diagnosed patients should be isolated from social networks to cut the transmission links (yellow edges). Furthermore, the red edges are more dangerous for the exposed persons and the comprehensive investigation policy for all people and the nucleic acid testing strategy for symptomatic patients should be taken to remove their connections with others. Meanwhile, some carrier would migrate to other place through the global transportation networks. The experiences from China demonstrated that the external input is the largest threat in the post-transmission period~\cite{chinazzi2020effect}.}
\end{figure}

\begin{figure}
  \centering
  \includegraphics[width=17cm]{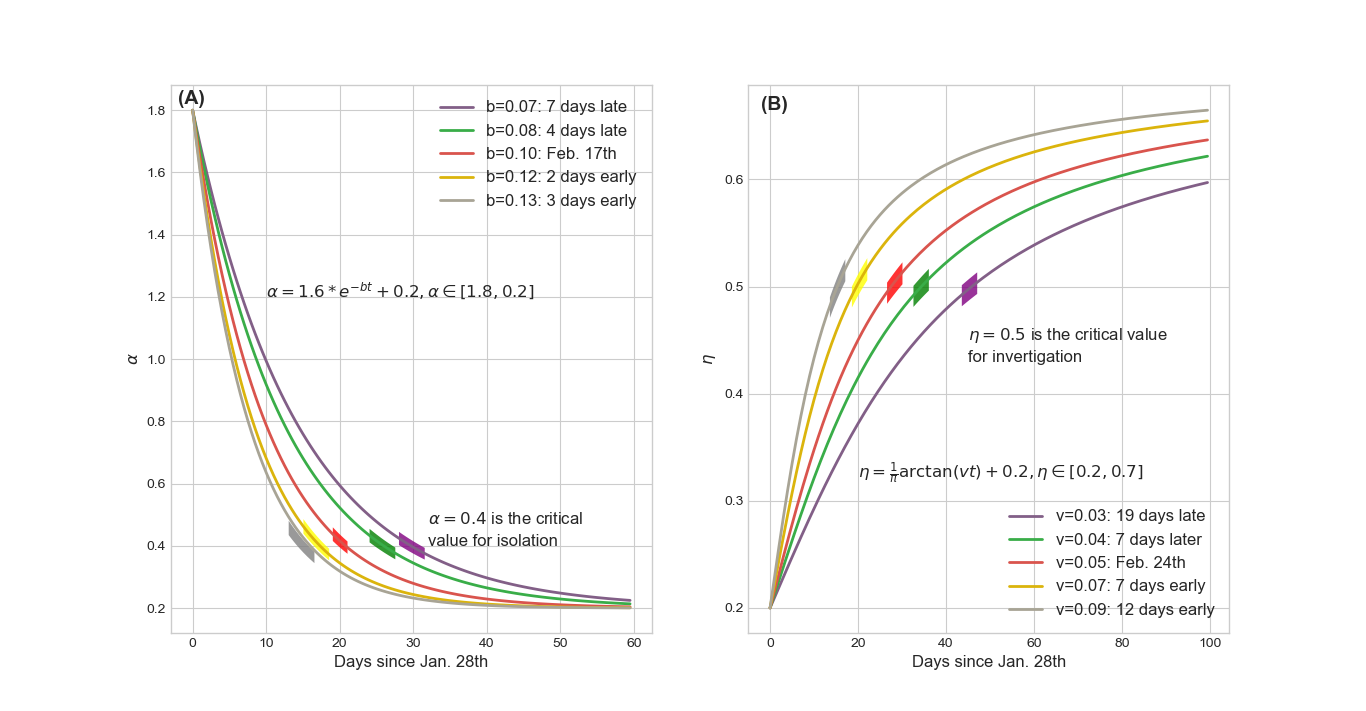}
  \caption{\textbf{Mathematical representations of dynamical coefficients with respect to isolation policy and investigation policy respectively.} \textbf{(A)} tuning the parameter $b$ of $\alpha$ can subtly regulate the implement time of isolation policy. We choose $\alpha=0.4$ as the critical value for isolation which this policy is fully implemented. The practical isolation policy curve $b = 0.10$ (Feb. 17th) is selected as the baseline to compare results caused by advancing or postponing of this policy. The boundaries of $\alpha \in [1.8, 0.2]$ are determined by the basic reproduction number $R_0$ and the duration from infected to infected. \textbf{(B)} tuning the parameter $v$ of $\eta$ can reflect the rate of community investigation for the infected but undiagnosed patients. We choose $\eta = 0.5$ as the threshold of daily investigation rate and $v = 0.05$ (Feb. 24th) as the baseline to compare the impact of the implement time of investigation policy to the pandemic trend. The boundaries of $\eta \in [0.2, 0.7]$ are determined by the case capacity of the hospitals.}
\end{figure}

\begin{figure}
  \centering
  \includegraphics[width=17cm]{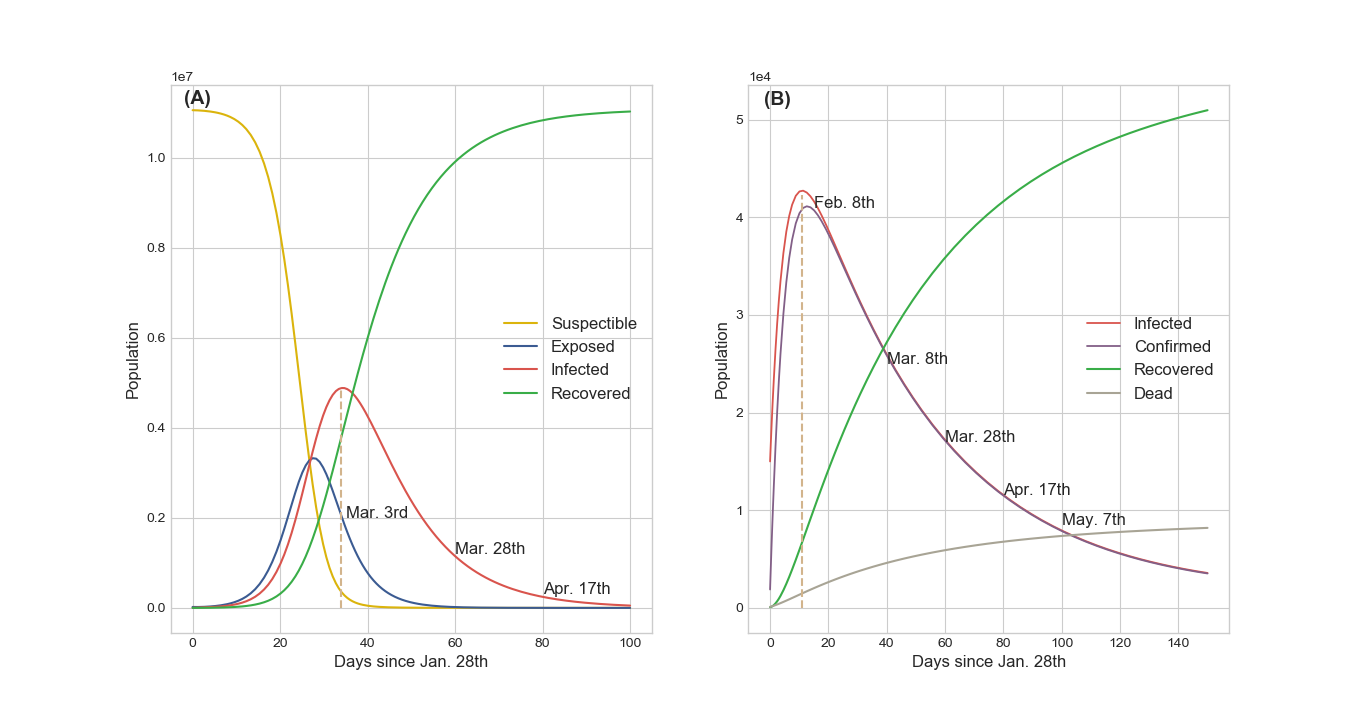}
  \caption{\textbf{Results from the numerical simulations about COVID-19 transmission tendency of SEIR and S-ICRD models.} \textbf{(A)} SEIR predicts that the existing infected cases would reach a peak of 4,885,672 on Mar. 3rd and the pandemic would be basically blocked on Apr. 17th in Wuhan. More than half of the population would be infected by COVID-19 but the fraction of exposed population is less than the infected population which is inconsistent with the observation. \textbf{(B)} S-ICRD predicts that the peak of the existing infected cases is 42,729 on Feb. 8th and the epidemic alert would be lifted after May. The forecasted number of infections is consistent with the actual data but the relative crucial time points and the final deaths have a great deviation with the observation in Wuhan since the constant coefficients are inappropriate to simulate the transmission dynamics of COVID-19.}
\end{figure}

\begin{figure}
  \centering
  \includegraphics[width=17cm]{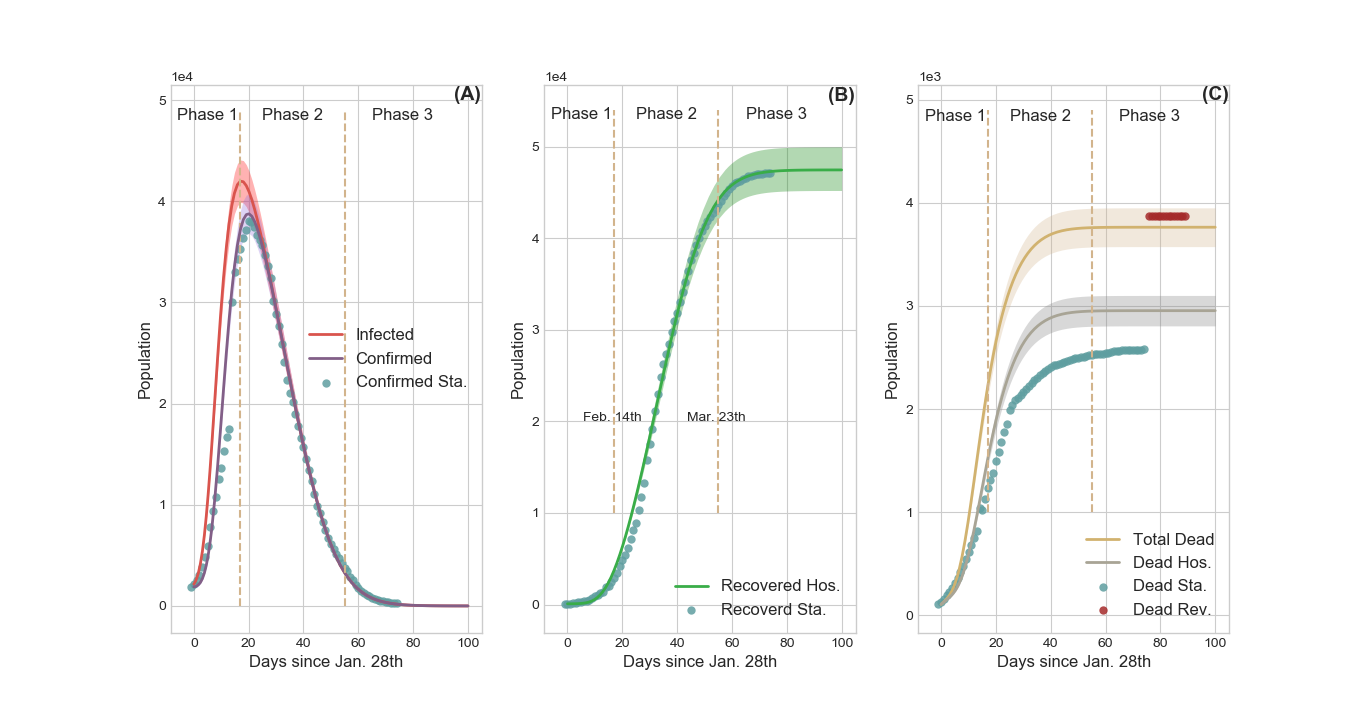}
  \caption{\textbf{Comparison of the results generated by D-ICRD model and the time series data released by NHC.} In this figure the dots denote official data and the lines denote the data calculated by D-ICRD in which the shadow areas denote the 95\% confidence interval (CI). The entire pandemic in Wuhan is divided into three phases: Phase 1 [Jan. 23th - Feb. 14th]; Phase 2 [Feb.15 - Mar. 23th]; Phase 3 [Mar. 24th - ]. Sta. denotes the official statistical data and Hos. denotes the data counted in hospitals. \textbf{(A)} the curves of the existing infected and confirmed cases. The gap between the infected curve and the confirmed curve indicates the number of the infected but unconfirmed group~\cite{mallapaty2020antibody}. Most dots are located in the shadow areas except for Phase 1 because the health authorities clarified that there were many infected cases of underreporting. \textbf{(B)} The curve of recovered cases. The calculated data is slightly larger than the amounts counted in hospitals since the spontaneously curative patients are not included and a few medical institutions failed to connect with the information systems in time. \textbf{(C)} The curve of death cases. Blue dots denote the originally released data and the red dots denote the revised data by NHC. According to the revised data, mortality is seriously underestimated at Phase 1 and Phase 2, and the yellow curve is more representative of the death trend of COVID-19 pandemic.}
\end{figure}

\begin{figure}
  \centering
  \includegraphics[width=17cm]{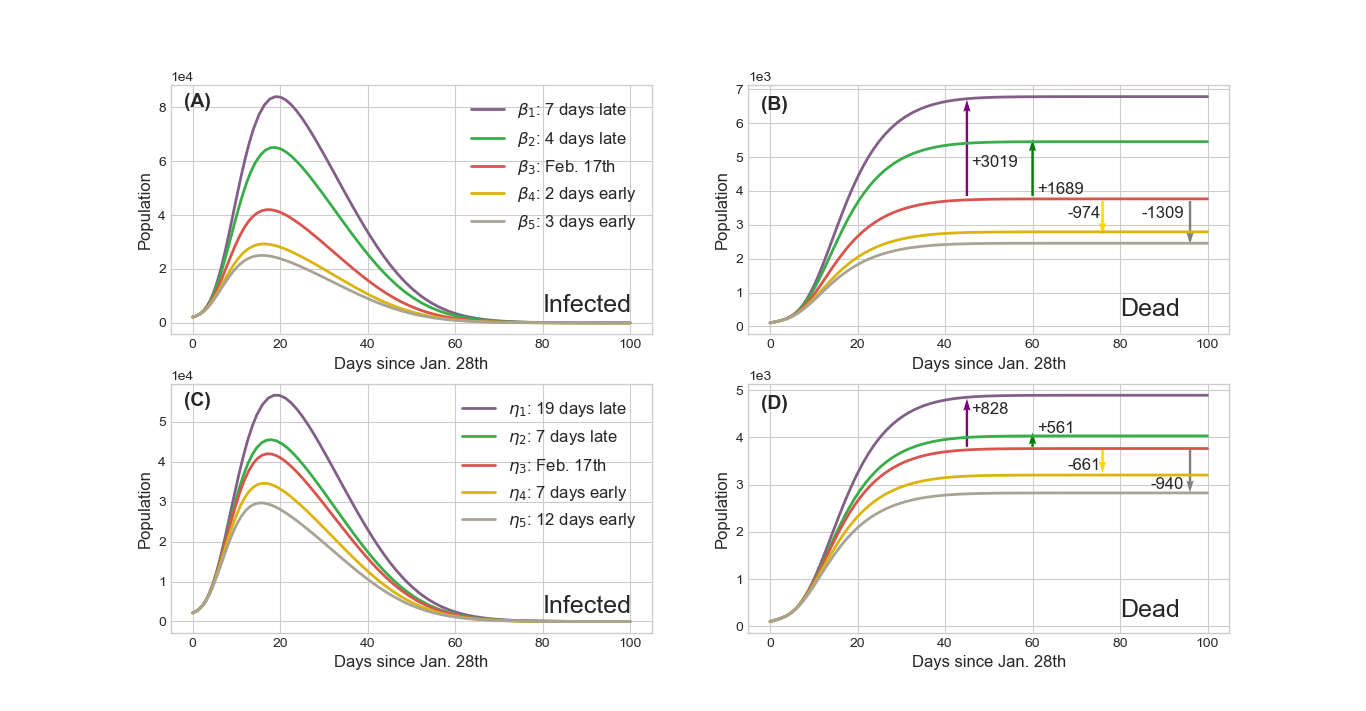}
  \caption{\textbf{Impact of the implement time of isolation and investigation policies on transmission tendency of COVID-19 in Wuhan.} The red lines are the results of the baseline policies selected in Fig. 2 and the other lines' colors are matched with Fig. 2 representing advance or delay the corresponding policy. The arrows show the increment or decrement of the death cases compared with the baseline of different policies. \textbf{(A)} and \textbf{(C)} The curves of infected persons calculated by the dynamical coefficient $\alpha$ (isolation policy) and $\eta$ (investigation policy) in Fig. 2 respectively. \textbf{(B)} and \textbf{(D)} The curves of deaths corresponding to \textbf{(A)} and \textbf{(C)} and they provide the statistic evidences how public health interventions save the lives of people.}
\end{figure}
\end{document}